# TRICRITICAL POINTS AND LIQUID-SOLID CRITICAL LINES

ANNELI AITTA

*Institute of Theoretical Geophysics, Department of Applied Mathematics and Theoretical Physics, University of Cambridge, Wilberforce Road Cambridge, CB3 0WA, UK*

Tricritical points separate continuous and discontinuous symmetry breaking transitions. They occur in a variety of physical systems and their mathematical models. A tricritical point is used to determine a liquid-solid phase transition line in the pressure-temperature plane [Aitta, J. Stat. Mech., 2006]. Excellent experimental agreement has been obtained for iron, the material having the most high pressure data. This allows extrapolation to much higher pressures and temperatures than available experimentally. One can predict the temperature at the liquid-solid boundary in the Earth's core where the pressure is 329 GPa. Light matter, present as impurities in the core fluid, is found to generate about a 600 K reduction of this temperature.

Melting or solidification is a first order phase transition since the order changes discontinuously from liquid to solid. Landau (1937) gave a theoretical description for first order phase transitions and the point where they change to second order phase transitions (with a continuous change of order) [1]. Such a point was later named a tricritical point by Griffiths (1970) [2]. Tricritical points occur in a variety of physical systems. Examples of tricritical points are presented in Table 1 with the corresponding adjustable variables. Experimentally they were first found in fluid mixtures, compressed single crystals and magnetic and ferroelectric systems (see old reviews in [3] and [4]). Paper [11] is an example of two-dimensional melting.

    The rest of this paper provides bifurcation theoretical analysis for the solidification/melting problem for iron [12], following the earlier work in Paper [13] which presents the symmetry breaking analysis in the first experimentally studied nonequilibrium tricritical point.

Table 1. Examples of tricritical points.

| Physical system | Variable 1 | Variable 2 |
|---|---|---|
| $^3$He-$^4$He mixtures [2] | Density or concentration | Temperature |
| Vortex-lattice melting [5] | Magnetic field | Temperature |
| Liquid crystals [6] | Concentration | Temperature |
| Cold Fermi gas [7] | Spin polarization | Temperature |
| Ferroelectrics [8] | Pressure or electric field | Temperature |
| Metamagnets [9] | Pressure or magnetic field | Temperature |
| Structural transition [10] | Pressure | Temperature |
| Melting on graphite [11] | Coverage | Temperature |
| Solidification [12] | Pressure | Temperature |
| Taylor-Couette vortex pair [13] | Aspect ratio | Rotation rate |

1. **Landau theory**

Following Landau [1], an order parameter $x$ can be used to describe first order phase transitions which change to be second order at a tricritical point. Here $x = 0$ for the more ordered solid phase which occurs at lower temperature, and in the less ordered liquid phase, $x \neq 0$. The Gibbs free energy density is proportional to the Landau potential, which needs to be a sixth order polynomial in $x$:

$$\Phi = x^6/6 + g x^4/4 + \varepsilon x^2/2 + \Phi_0. \qquad (1)$$

A set of examples of $\Phi - \Phi_0$ is shown in the Fig. 1. No higher order terms in $x$ appear in $\Phi$ since they can be eliminated using coordinate transformations as in bifurcation theory [14]. This method also scales out any dependence on physical parameters of the coefficient of the $x^6$ term. Generally $\Phi_0$, $\varepsilon$ and $g$ depend on the physical parameters and for solidification they are pressure $P$ and temperature

*T*. In equilibrium, the order parameter takes a value where the potential $\Phi$ has a local or global minimum. The minima of $\Phi$, the solutions of

$$x^5 + g x^3 + \varepsilon x = 0 \tag{2}$$

at which $d^2\Phi/dx^2$ is positive, give three stable equilibrium states provided $0 < \varepsilon < g^2/4$ and $g<0$. They are at

$$x = 0 \text{ and } x = \pm\sqrt{-g/2 + \sqrt{g^2/4 - \varepsilon}}. \tag{3}$$

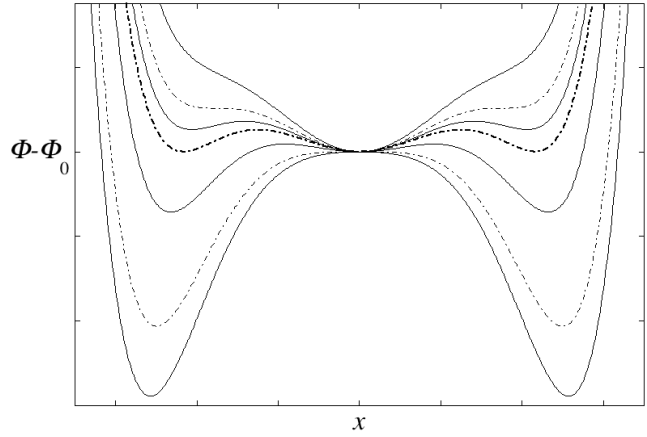

Figure 1. Order parameter *x* dependent part of the Landau potential (1) for a fixed *g*<0 and various values of *ε*.

The thermodynamic transition from liquid to solid occurs when all three minima of the potential are equally deep: $\Phi - \Phi_0 = 0$ at those three values of *x*. This happens when

$$\varepsilon = 3g^2/16 \tag{4}$$

corresponding to the middle dash-dotted curve in Fig. 1. The liquid phase is then in thermal equilibrium with the solid phase. If $\varepsilon > 3g^2/16$ the solid is preferred, and if $\varepsilon < 3g^2/16$ the liquid. There are also two other critical conditions: Liquid phase exists as an unfavoured state until the potential changes from having three minima to one minimum (the highest dash-dotted curve in Fig. 1), that is, at

$$\varepsilon = g^2/4. \tag{5}$$

The solid phase exists as an unfavoured state until $\varepsilon = 0$ where the potential changes from having three to two minima (the lowest dash-dotted curve in Fig. 1).

These liquid-solid phase transitions can be presented using simple bifurcation diagrams as in Fig. 2. In the direction where $T$ increases (see Fig. 2a), for $T<T_S$ the solid state (having $x=0$) is the only possible state of the system. At higher temperatures solid state is preferred but liquid state is possible as an unfavoured stable state until $T=T_M$ where the melting occurs. At higher temperatures the liquid state is preferred but solid state can occur as an unfavoured stable state until $T=T_L$ beyond which only liquid exists because the solid state is unstable. Now consider the direction where $P$ increases (see Fig. 2b which has $T>T_L$). For $P<P_L$ only liquid state is stable. There solid state is unstable and thus not a possible state of the system. At $P=P_L$ the unstable solid state bifurcates to a stable solid state and two unstable liquid states (only the branch with positive $x$ is shown) which turn backward to be stable states at $P=P_S$. For $P_L<P<P_M$ liquid state is still preferred but solid state is possible as an unfavoured stable state. At $P=P_M$ solidification occurs. At higher pressures solid state is preferred but liquid state can occur as an unfavoured stable state until $P_S$ beyond which only solid state is stable and thus the only possible state of the system.

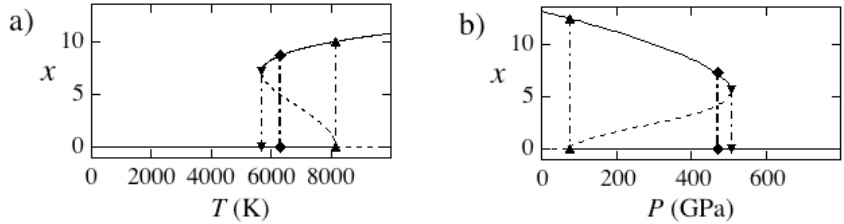

Figure 2. Bifurcation diagrams for non-negative orderparameter $x$ as function of (a) temperature and (b) pressure. Symbols show the values of the critical conditions: downward triangle, diamond and triangle correspond to subindex S, M and L, respectively.

These simple backward bifurcations in two separate physical parameter directions can be expressed in a combined way by using bifurcation theory. The relevant normal form (see Table 5.1, form (8) in Ref. [14]) for this tricritical bifurcation is $H_0 = x^5 + 2m\lambda x^3 - \lambda^2 x$ and its universal unfolding is

$$H = x^5 + 2m\lambda x^3 - \lambda^2 x + \alpha x + \beta x^3 \tag{6}$$

where $\lambda$ is the bifurcation parameter, $\alpha$ and $\beta$ are the unfolding parameters and $m$ the modal parameter. Bifurcations are assumed to be perfect. The equation $H=0$ is equivalent to Eq. (2) if one identifies

$$g = 2m\,\lambda + \beta \qquad (7)$$

and

$$\varepsilon = -\lambda^2 + \alpha. \qquad (8)$$

For some materials the melting curve in the ($P$,$T$) plane is expected to have a horizontal tangent at high $P$. Here the starting point of this tangent is identified as the tricritical point ($P_{tc}$,$T_{tc}$). Thus $g$ axis in the Landau theory can be identified to be parallel to $P$ axis. In Ref. [12] $\varepsilon$ axis was assumed to be parallel to $T$ axis. Here $\varepsilon$ is allowed to depend on both $T$ and $P$. This dependence can be found by taking the critical temperatures to depend quadratically on $P$ since that is the highest power relationship between $\varepsilon$ and $g$ on the critical lines in the Landau theory. Then all three critical lines can be expressed as

$$T_{tc} - T = a_i\,(P - P_{tc})^2, i = 1, 2, 3. \qquad (9)$$

When $i=1$ we have the curve where $\varepsilon=0$ and $T=T_L(P)$. Denoting by $T_{L0}$ the value of $T_L$ at $P = 0$, one finds $a_1$ so that

$$T_{tc} - T_L(P) = (T_{tc} - T_{L0})(P/P_{tc} - 1)^2. \qquad (10)$$

Moving all the terms to one side allows one to write generally

$$\varepsilon = T_{tc} - T - (T_{tc} - T_{L0})(P/P_{tc} - 1)^2 \qquad (11)$$

which is in the form (8) if $\alpha = T_{tc} - T$ and

$$\lambda = \sqrt{T_{tc} - T_{L0}}\,(P/P_{tc} - 1) \qquad (12)$$

since $\lambda<0$. Now one can simplify (7) to

$$g = 2m\,\lambda \qquad (13)$$

since $g=0$ at $P=P_{tc}$. For first order transitions $g$ needs to be negative. So for $\lambda$ as above, $m$ needs to be positive.

When $i=2$ we have the melting curve: $T=T_M(P)$. At $P=0$, $T_M=T_0$ gives $a_2$. Thus the equation of the melting curve is

$$T_M(P) = T_{tc} - (T_{tc} - T_0)(P/P_{tc} - 1)^2. \tag{14}$$

Inserting this in Eq. (11) one obtains

$$\varepsilon_M = (T_{L0} - T_0)(P/P_{tc} - 1)^2. \tag{15}$$

Combining this with (4) and (13) one can find

$$m = 2\sqrt{(T_{L0} - T_0)/[3(T_{tc} - T_{L0})]} \tag{16}$$

When $i=3$ we have the third critical curve marking the end of the hysteresis, the condition for the lowest possible liquid phase temperature $T_S$. The liquid state must vanish at $T=0$, so at $P=0$ $T_S=0$ gives $a_3 = T_{tc}/P_{tc}^2$. Thus the equation of the hysteresis curve is

$$T_S(P) = T_{tc} - T_{tc}(P/P_{tc} - 1)^2 \tag{17}$$

and inserting this in (11) one finds

$$\varepsilon_S = T_{L0}(P/P_{tc} - 1)^2. \tag{18}$$

Using this with (5) one obtains $T_{L0} = 4T_0$ allowing to write the equation (10) as

$$T_L(P) = T_{tc} - (T_{tc} - 4T_0)(P/P_{tc} - 1)^2. \tag{19}$$

Thus one finds

$$g = 4\sqrt{T_0}(P/P_{tc} - 1) \tag{20}$$

and

$$\varepsilon = T_{tc} - T - (T_{tc} - 4T_0)(P/P_{tc} - 1)^2 \tag{21}$$

and

$$x = 0 \text{ or } x = \pm\sqrt{2\sqrt{T_0}(1 - P/P_{tc}) \pm \sqrt{T - T_{tc} + T_{tc}(P/P_{tc} - 1)^2}}. \tag{22}$$

These values of $x$ are drawn in Fig. 2 using the iron tricritical point obtained from the experimental data as discussed next. In Fig. 2a $x(T)$ is shown for $P=329$ GPa (corresponding to the pressure on Earth's inner core boundary [15] which is a solidification front in iron-rich core melt). In Fig. 2b the bifurcation structure of $x(P)$ is shown for $T=7500$ K which is greater than $T_{L0}$, thus exhibiting all three critical transitions. The critical curves (14), (17) and (19) are drawn in Fig. 3 with iron melting data and *ab initio* calculations.

## 2. Experimental evidence for iron

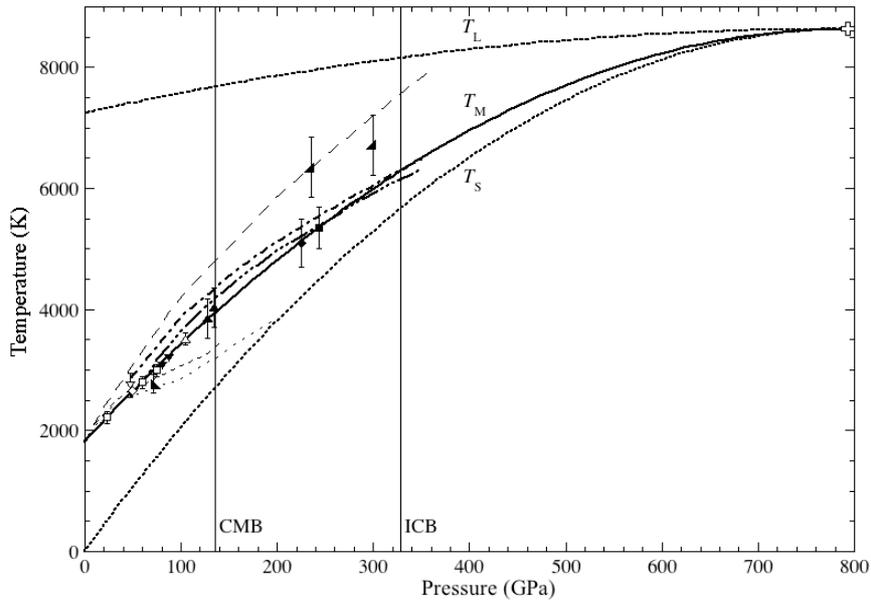

Figure 3. Iron melting experimental data, with *ab initio* calculations without (dash-dotted) and with (dash-triple-dotted) a free energy correction [16] and three theoretical critical lines $T_S$, $T_M$ and $T_L$ from this work. The most reliable static data (open symbols: squares [17], diamond [18] and triangle [19]) with consistent shock wave data (filled symbols: square [20], diamond [21], triangles [22] and inverted triangles [23]) are used to find the theoretical curve for $T_M$ (solid line) and thus the coordinates of the tricritical point ($P_{tc}$, $T_{tc}$) (cross). Other data (see discussion in [12]) are also shown: short dashed [24] and dashed [25] lines and open inverted triangle [26] are for static data, filled symbols are for shock data (lower right corner triangles [27], lower left corner triangle [28]), but the long dashed line [29] represents static data extrapolated to high pressures using shock data. The vertical lines show the pressures at the Earth's core mantle boundary (CMB) and inner core boundary (ICB).

Iron is the dominant element in terrestrial planetary cores. For instance, the Earth has an iron-rich core at depths below about half of the Earth's radius. The outer core is molten but the inner core is close to pure iron which is solidifying out from the outer core melt. Owing to its significant geophysical interest, iron is the most studied high pressure material. In Fig. 3, data since 1986 is presented with the *ab initio* calculations for iron melting. For discussion, see [12]. Overall, the data has a large scatter, and all shock wave results as well as the older static measurements have very large error bars. However, selecting (details in Ref. [12]) the most reliable static results combined with all supporting high pressure shock results allows one to find an excellent fit to Eq. (14) giving the tricritical point as (793 GPa, 8632 K) with a correlated uncertainty of about ± (100 GPa, 800 K). The theoretical melting curve (14) is drawn in Fig. 3 and it goes approximately through the middle of the data. In addition, the curves for $T_L$ from Eq. (19) and $T_S$ from Eq. (17) are also drawn showing the limits of the range where unfavoured liquid or solid stable states can occur. All the experimental data are in this range.

The pressure at the Earth's core-mantle boundary (CMB) is about 136 GPa [15]. The iron melting temperature is at that pressure 3945 ± 12 K using Eq. (14). This temperature is very similar to the seismic estimate 3950±200 K [30] for temperature there implying the melt is rather close to pure iron. This agrees very well with the result that experimentally, pure iron compressional wave velocities are very similar to seismic velocities in the Earth's core melt close to the CMB (see fig. 2 in Ref. [31]). However, the mantle solid at the CMB has density of about 56 % of the core melt density there [15]. Thus it is the light matter in the core melt which is solidifying out there, presumably with some iron. This solidification temperature stays very close to pure iron for small concentrations of light matter in the iron-rich melt. Solidification at CMB has been suggested previously [32-34].

The iron melting temperature at the pressure of the Earth's inner core boundary (ICB) is 6290 ± 80 K. This is very close to the *ab initio* result without free energy correction (see Fig. 3). However, the inner core is solidifying from the molten iron-rich outer core, which owing to the seismic density estimates [15] there is concluded to have also some light elements. These light impurities are lowering the temperature from the pure iron melting temperature. From this work we can conclude from Eq. (17) that in the real Earth the temperature at the ICB is possible to be as low as $T_S$, about 5670 K, but not lower. Since the fluid in the outer core has been convecting for billions of years it has been able to adjust its fluid concentration and temperature profiles so that the temperature and density gradients inside the fluid are minimized. Thus the temperature at the ICB is expected to take this limiting value of about 5670 K. This estimate agrees very well with the value of 5700 K as inferred for the temperature at the ICB from *ab initio* calculations on the elasticity properties of the inner core [35] and is consistent with the range 5400 K – 5700 K reported in Ref. [36]. The temperature difference $T_M - T_S$ = 621 K at ICB also agrees with the estimate 600 K to 700 K in Ref. [36] but is more than twice the 300 K used in the rather recent energy budget calculations of the core [37].

## 3. Conclusions

The concept of tricritical point seems to be very useful in considering liquid-solid phase transitions as a function of pressure and temperature. Landau theory gives a quadratic, general formula for the solidification/melting curve $T_M(P)$ if it ends at a tricritical point where $dT_M/dP=0$. The structure of the coefficients is definite, but the tricritical point needs to be estimated from good solidification or melting data. For pure iron, the temperature formula and data agree very well in the whole range 0–250 GPa where we have good experimental data. The tricritical point is estimated to be at (800±100 GPa, 8600±800 K), with the signs of the errors correlated. The prediction for the iron melting temperature at the core-mantle boundary is 3945±12 K, very similar to the seismic estimate for temperature of 3950±200 K at the CMB implying the melt there is rather close to pure iron. The impurities present in the outer core decrease the inner core boundary temperature from 6290±80 K found for pure iron to about 5670 K for the real Earth, in agreement with *ab initio* predictions.